\documentclass[prl,twocolumn,superscriptaddress]{revtex4-1}

\usepackage{amsmath}
\usepackage{amssymb}
\usepackage{amsthm}

\usepackage{graphicx}
\usepackage{graphics}
\usepackage{bbold,bm}
\usepackage{hyperref}
\usepackage{epsfig}
\usepackage{dcolumn}
\def\FMF{Floquet Majorana fermion}

\def\ket#1{\mathinner{|{#1}\rangle}}
\def\braket#1{\mathinner{\langle{#1}\rangle}}

\def\bbraket#1{\mathinner{\langle\hspace{-0.75mm}\langle{#1}\rangle\hspace{-0.75mm}\rangle}}

\def\dd{\mathrm{d}}

\def\ii{i}
 
\def\dbar{\hbox{$d$\kern-0.6em\raise0.3em\hbox{$-$}}\hspace{-0.5mm}}

\begin{document}

\title{Tunable Floquet Majorana Fermions in Driven Coupled Quantum Dots}

\author{Yantao Li}
\affiliation{State Key Laboratory of Optoelectronic Materials and Technologies, School of
Physics and Engineering, Sun Yat-sen University, Guangzhou 510275, People's
Republic of China}
\affiliation{Department of Physics, Indiana University, 727 East Third Street, Bloomington, IN 47405-7105 USA}

\author{Arijit Kundu}
\affiliation{Department of Physics, Indiana University, 727 East Third Street, Bloomington, IN 47405-7105 USA}

\author{Fan Zhong}
\affiliation{State Key Laboratory of Optoelectronic Materials and Technologies, School of
Physics and Engineering, Sun Yat-sen University, Guangzhou 510275, People's
Republic of China}

\author{Babak Seradjeh}
\affiliation{Department of Physics, Indiana University, 727 East Third Street, Bloomington, IN 47405-7105 USA}

\begin{abstract}
We propose a system of coupled quantum dots in proximity to a superconductor and driven by separate ac potentials to realize and detect Floquet Majorana fermions. We show that the appearance of Floquet Majorana fermions can be finely controlled in the expanded parameter space of the drive frequency, amplitude, and phase difference across the two dots. While these Majorana fermions are not topologically protected, the highly tunable setup provides a realistic system for observing the exotic physics associated with Majorana fermions as well as their dynamical generation and manipulation.
\end{abstract}

\maketitle

\emph{Introduction.}---Majorana fermions are spin one-half particles that are their own anti-particles. Their existence as elementary particles in nature is speculated in several theories, though conclusive experimental evidence is still lacking~\cite{Wil09a,AugAutBar12a}. They can also exist as collective quasiparticles in condensed matter systems, in particular as bound states at the edges or vortices of a topological superconductor~\cite{Ali12a,Bee13a}. In addition to their interesting physical properties, Majorana bound states in a superconductor encode non-Abelian exchange statistics.  More recently, Majorana fermions have also been theoretically found to exist as steady states in a periodically driven system even under conditions for which the instantaneous static system would not host them anywhere in the range of the drive~\cite{JiaKitAli11a,LiuHaoZhu12a,LiuLevBar13a,ReyFru13a,KunSer13a}. The exchange statistics of these ``Floquet Majorana fermions'' are predicted to be similar to their static counterparts, thus providing a new avenue for applications in fault-tolerant quantum information processing~\cite{Iva01a,Kit03a}. Moreover, \FMF s have a richer topological classification~\cite{RudLinBer13a} and open a new frontier of our understanding of nonequilibrium quantum systems. 

Though preliminary signatures of the existence of static Majorana fermions have been reported in several experiments~\cite{MouZuoFro12a,DasRonMos12b,WilBesGal12b,RokLiuFur12a,FinVanMoh13a}, the results have been subject to  interpretation~\cite{LiuPotLaw12b,PikDahWim12b} in part due to the ambiguity of the mechanism leading to the observed signatures and the complexity of the physical setup. A simpler setup using quantum dots has been proposed and studied by several authors~\cite{Fle11a,LeiFle12a,SauSar12a}. In the simplest case of just two quantum dots, Majorana fermions lose their topological protection. However, the advantages of unambiguous detection and simpler design favor these proposals in the lab setting. In the absence of topological protection, the system must be finely tuned by a rather intricate use of variable magnetic fields at the nanoscale. 

\begin{figure}
\centering
\includegraphics[width=0.43\textwidth]{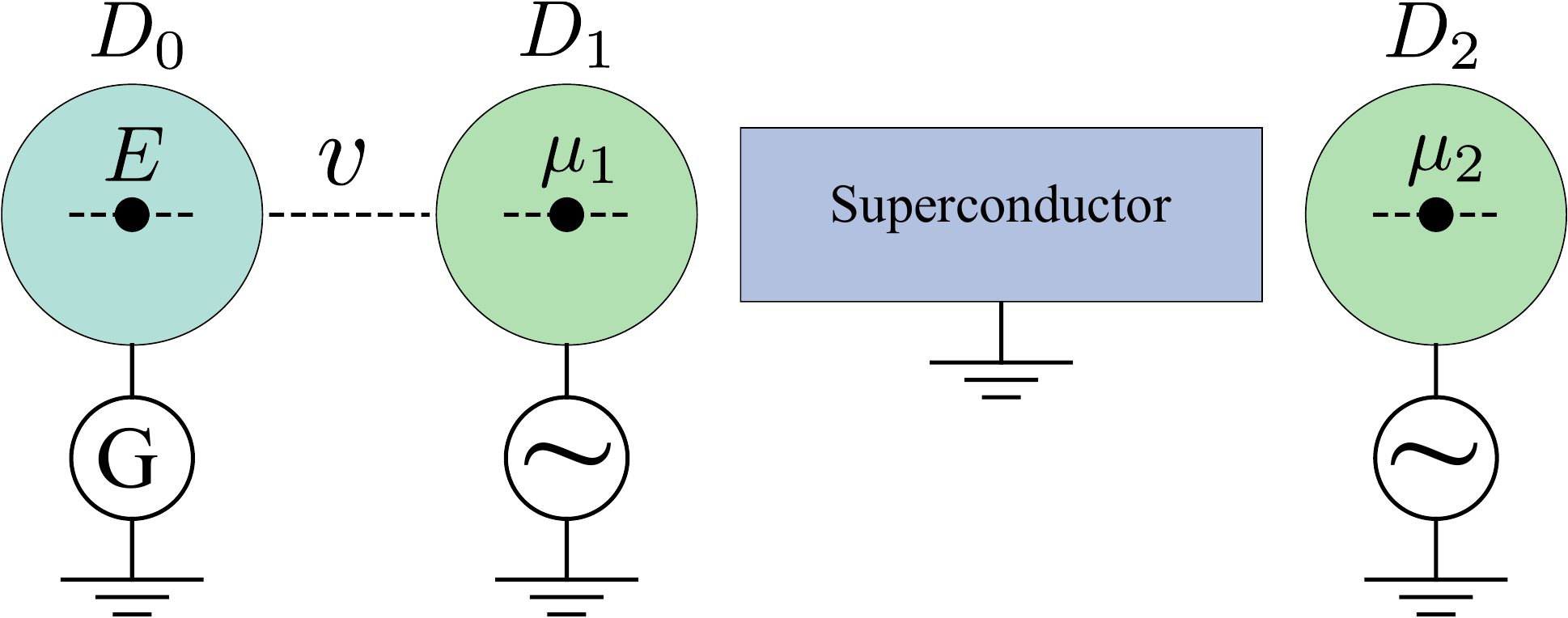}
\caption{(color online) The schematic of the proposed setup. Two dots, $D_1$ and $D_2$, each with a single level at $\mu_1$ and $\mu_2$, respectively, are coupled by proixmity to a superconductor. An external ac voltage is applied to $D_1$ and $D_2$ to realize steady-state Floquet Majorana fermions. A probe dot, $D_0$, is coupled to the other dots with tunneling amplitude $v$ and used to detect the Floquet Majorana fermions hosted by $D_1$ and $D_2$ by varying the energy $E$ of the resonant level of $D_0$ through gate $G$.}
\label{fig:setup}
\end{figure}

In this paper, we propose to overcome these shortcomings by periodically driving the quantum dots externally. The schematic setup of our proposal is shown in Fig.~\ref{fig:setup}. The quantum dots are proximity-coupled to a superconductor which induces Cooper pair correlations across the dots. For simplicity of our analysis, we assume that the lowest state in each dot is well separated from the rest and is non-degenerate. (The experimental conditions for realizing such a system are discussed later.)  Steady state \FMF s are generated by the external drive instead of a variable magnetic field, which also adds several highly tunable experimental knobs to address the fine-tuning requirement. \FMF s are detected unambiguously by measuring the charge of a probe dot as its energy level is varied adiabatically via a gate potential. Even though there is no topological protection in this simplest case of two quantum dots, the exquisite tunability of the setup and a clear detection signal provide a realistic and viable system for observing the physics of Majorana fermions, on the one hand, and a venue to study the the physics of dynamical topological bound states, on the other hand. 

\emph{Model.}---The effective Hamiltonian of the system is $H(t)=H_0+V(t)$. The static part
\begin{equation}
H_0 = \sum_i\mu_{i}d_{i}^{\dagger}d_{i}+\left( \lambda d_{1}^{\dagger}d_{2}+\Delta d_{1}^{\dagger}d_{2}^{\dagger} + \mathrm{h.c.} \right),
\label{eq:Ham0}
\end{equation}
where $d_{i}^{\dagger}$ and $d_{i}$ are the creation and annihilation operators of the state of dot $i=1,2$ with energy $\mu_{i}$ measured with respect to the chemical potential of the superconductor, $\lambda$ is the effective hopping amplitude and $\Delta$ is the effective pairing amplitude between the dots. Due to charge screening by proximity to the superconductor we neglect any inter-dot Coulomb interaction, whereas intra-dot Coulomb interaction is irrelevant as only one effective energy level is available at each dot. Additionally, the dots are also subject to ac potentials,
\begin{align}
V(t) =A_0\cos(\Omega t) d_{1}^{\dagger}d_{1}+ A_0 \cos(\Omega t+\theta) d_{2}^{\dagger}d_{2},\label{driven}%
\end{align}
where $A_0$ is the driving amplitude, $\Omega$ the frequency and $\theta$ the phase difference between the ac potentials. The amplitude of the two driving potentials are assumed to be the same for simplicity. We have checked numerically that different amplitudes do not change our conclusions.



\begin{figure}[t]
\begin{center}
\includegraphics[width=0.4\textwidth]{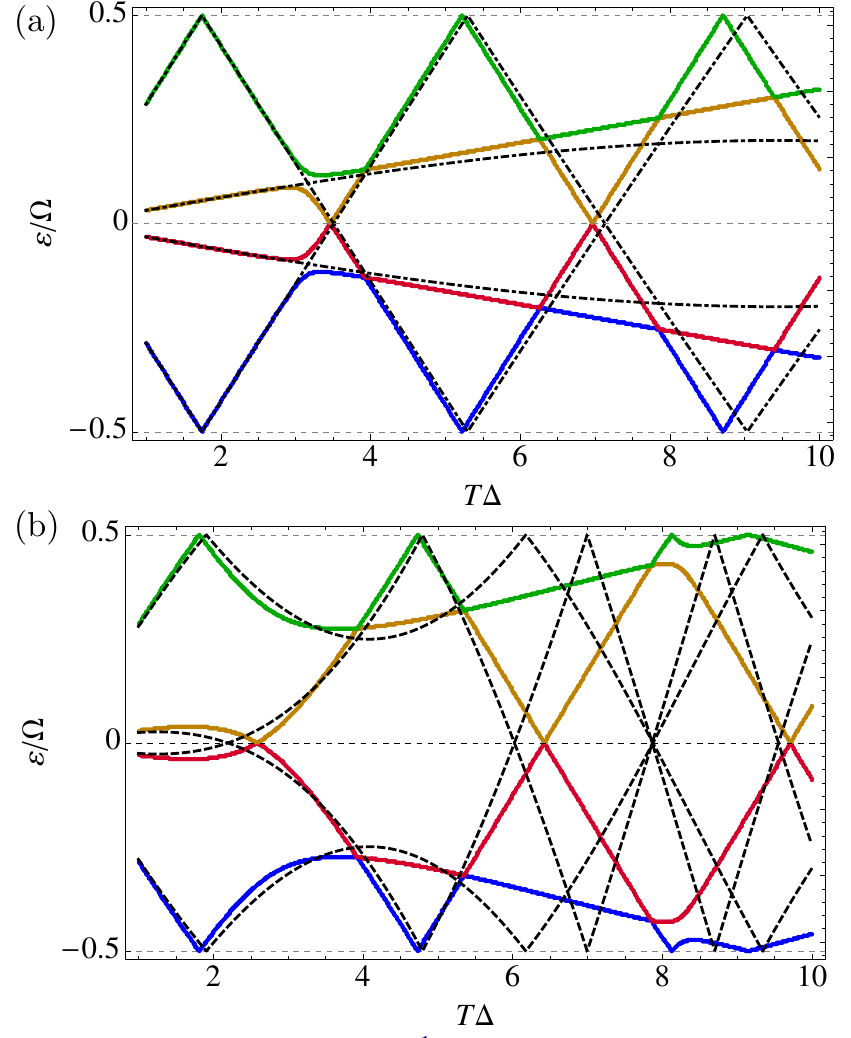}
\end{center}
\caption{(color online) The appearance of Floquet Majorana fermions. Quasienergies $\varepsilon$ of the coupled-dot system is shown as a function of period $T=2\pi/\Omega$ of the ac voltage $A_0\cos(\Omega t)$ on both dots. Here $\mu_1=\mu_2=0$, the ratio of electron hopping to superconducting amplitude $\lambda/\Delta=0.8$, and (a) $A_0/\Delta=0.1$ and (b) $A_0/\Delta=0.75$. The solid lines are exact numerical solutions. The dashed lines are the approximate solutions of the effective Hamiltonian in the large frequency regime $\varepsilon\ll\Omega$. The Floquet Majorana fermions appear at quasienergies $0$ and $\pm\Omega/2$. The effective Hamiltonian approximation works well for small $A_0\lesssim\Omega$.}
\label{fig:band}
\end{figure}

\emph{Floquet Majorana fermions.}---When the system is driven with period $T\equiv2\pi/\Omega$, owing to the discrete time translation symmetry, the solutions of the Schr\"odinger equation can be written as $\ket{\psi_{\alpha}(t)} = e^{-\ii\varepsilon_{\alpha}t}\ket{\phi_{\alpha}(t)}$ with periodic Floquet wavefunction $\ket{\phi_{\alpha}(t+T)}=\ket{\phi_{\alpha}(t)}$ and quasienergy $\varepsilon_\alpha\in(-\Omega/2,\Omega/2]$, respectively, the eigenvector and eigenvalue of the Sch\"odinger-Floquet equation $[H(t)-i\partial_t]\ket{\phi_{\alpha}(t)}=\varepsilon_{\alpha}\ket{\phi_{\alpha}(t)}$ ($\hbar=1$ throughout). Accordingly, an extended Hilbert space may be defined~\cite{Sam73a} with the inner product $\bbraket{\phi|\phi'}=(1/T)\int_0^T\braket{\phi(t)|\phi'(t)}\dd t$. The Floquet wavefunctions and the quasienergies can be computed by solving the eigensystem
\begin{align}
U(T,0)\ket{\phi_{\alpha}\left(0\right)}=\exp\left(  -i\varepsilon_{\alpha}T\right) \ket{\phi_{\alpha}\left(0\right)},\label{eq:U}
\end{align}
where $U(t,t')$ is the time evolution operator $U(t,t')\ket{\psi(t')}=\ket{\psi(t)}$. The particle-hole symmetry of the Hamiltonian, $H\mapsto-H$ when $d_i\mapsto d_i^\dagger$, requires the quasienergies to come in pairs $(\varepsilon_\alpha,-\varepsilon_\alpha)$. \FMF s are states with quasienergy $\varepsilon_0=0$ or $\varepsilon_\pi=\pi/T=\Omega/2$~\cite{JiaKitAli11a}.

\emph{Large frequency approximation}---We may expand the periodic Hamiltonian and Floquet wavefunction in Fourier series $H(t)=\sum_{m}e^{im\Omega t}H^{(m)}$ and $|\phi_\alpha(t)\rangle=\sum_me^{im\Omega t}|\phi_\alpha^{(m)}\rangle$ and approximate the Schr\"odinger-Floquet equation as $H_{\mathrm{eff}}|\phi_\alpha^{(0)}\rangle = \varepsilon_\alpha|\phi_\alpha^{(0)}\rangle$, where the effective Hamiltonian~\cite{KitOkaBra11a}
\begin{align}
 H_{\mathrm{eff}} = H^{(0)} + \frac{\left[H^{(-1)},H^{(1)}\right]}{\Omega}+\mathcal{O}(\varepsilon/\Omega)^2.\label{eq:effH}
\end{align}
It is useful to use the \textit{rotating wave} basis
\begin{align}
\ket{\{n_i\},m}=\exp\left[  im\Omega t-i\int_0^tV(s) \dd s\right]  \ket{\{n_i\}}, \label{RWA}%
\end{align}
where $\ket{\{n_{i}\}}$ are the energy eigenstates of the static system $H_0$ with $n_i$ occupation in the dot $i$ and $m\in\mathbb{Z}$ labels the Fourier component. In this basis $H_{\mathrm{eff}}$ has the same form as $H_0$ with $\lambda$ and $\Delta$ renormalized to
\begin{align}
 \tilde{\lambda} =\lambda \, f\left(\frac{2A_0}{\Omega}\sin\frac\theta2\right), ~
  \tilde{\Delta} = \Delta\, f\left(\frac{2A_0}{\Omega}\cos\frac\theta2\right), 
\end{align}
where $f(x) = J_0(x)- x J_1(x)$.
The details of this calculation are given in the Supplemental Material~\cite{Note1}. 

\emph{Numerics.}---In Fig.~\ref{fig:band}, we show typical quasienergy spectra of the system. \FMF s with quasienergies $\varepsilon_0$ and $\varepsilon_\pi$ appear frequently at non-universal, parameter-dependent values of frequency. We also compare the quasienergies obtained in large frequency approximation with the exact numerical result in Fig.~\ref{fig:band}. For $A_0/\Omega\lesssim 1$ we obtain a reasonably good agreement with the exact numerical diagonalization of the Schr\"odinger-Floquet equation. We note that the second term in~(\ref{eq:effH}) results in a significant improvement of the approximation, essentially producing the first few \FMF s for both $\varepsilon_0$ and $\varepsilon_\pi$. 


 \begin{figure}[t]
\centering
\includegraphics[width=0.48\textwidth]{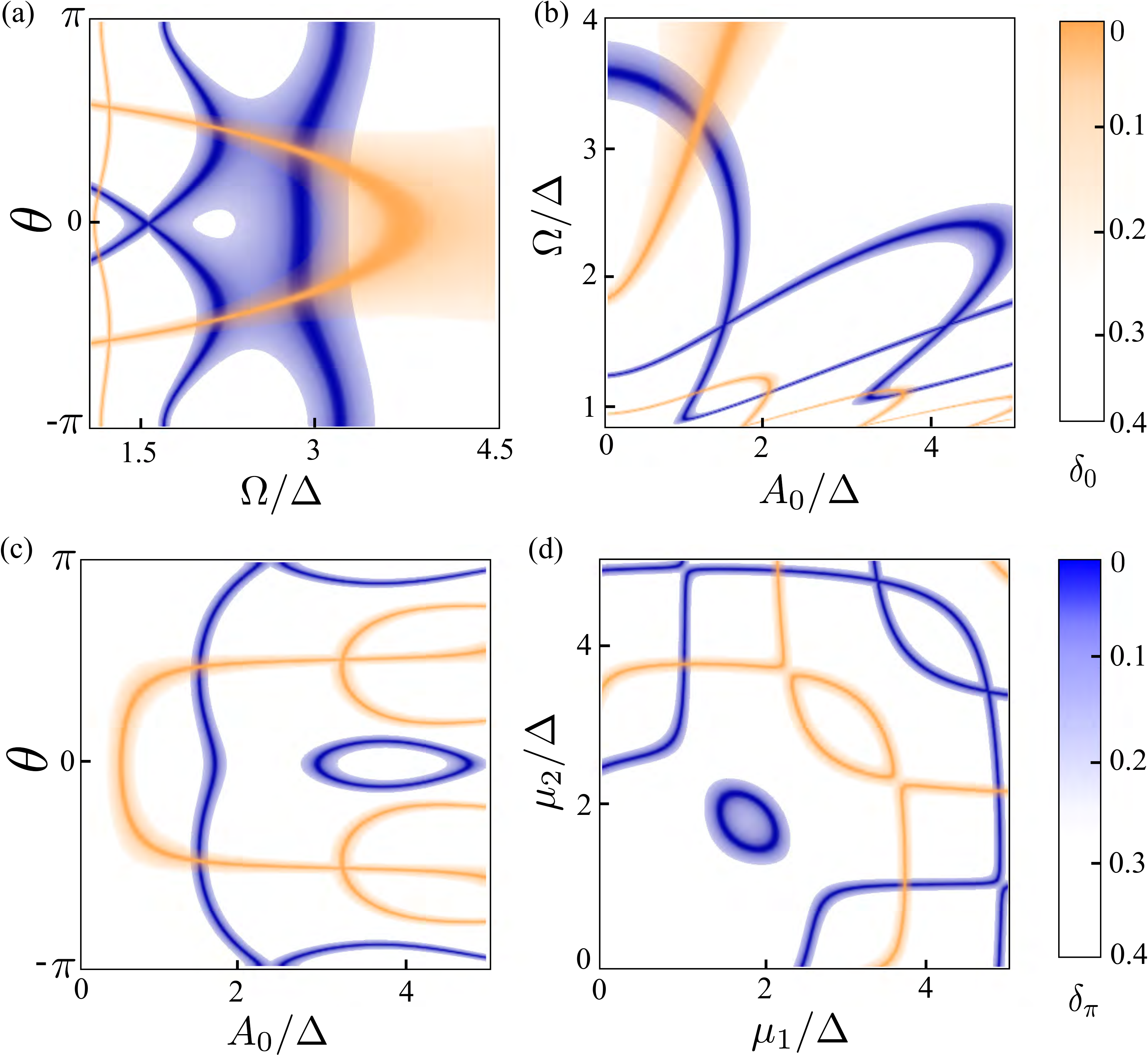}
\caption{(color online) Tuning \FMF s in the parameter space. The regions of existence of \FMF s are shown for ac voltages $A_0\cos(\Omega t)$ and $A_0\cos(\Omega t+\theta)$ on the two dots. The light (orange) and dark (blue) shades show, respectively, the quasienergy gaps $\delta_{0} = \varepsilon_1T$ and $\delta_{\pi} = \pi - \varepsilon_2T$ where $0<\varepsilon_1<\varepsilon_2$ are the quasienergies. Whenever fixed, the parameters are $\lambda/\Delta=0.8$, $\mu_1=\mu_2=0$, $\theta=0$, $A_0/\Delta=1.5$ and $\Omega/\Delta=2.1$. The system can be tuned easily to host \FMF s with quasienergy $0$ (orange), $\Omega/2$ (blue) or both. The thicker the shaded area, the more robust \FMF s are.}
\label{fig:phase}
\end{figure}

\emph{Fine-tuning and protection.}---The \FMF s in this system lack topological protection against perturbations due to mixing between the two dots and require fine-tuning to exist. The fine-tuning problem is addressed here by a large, controllable parameter space. If either $\mu_1=0$ or $\mu_2=0$ the condition to host $\varepsilon_0$ \FMF s can be found~\cite{LeiFle12a} from the zero energy states of $H_0$, i.e. $|\tilde\lambda| = |\tilde\Delta|$. It can therefore be seen from the shape of $f$ that for $A_0/\Omega\sim 1$ varying $\theta$ provides a very effective way of fine-tuning the system to host \FMF s.

In Fig.~\ref{fig:phase}, we depict the occurrence of \FMF s at $\varepsilon_0$ and $\varepsilon_\pi$ from the exact numerical solution in different planes of the parameter space. Several remarks are in order here: (1) \FMF s  appear in an extended region of the parameter space; (2) Varying the system parameters in a reasonably physical range produces many \FMF\ states, thus allowing the system to be fine-tuned to host \FMF s. In an experiment, both $A_0$ and $\theta$ can be varied with a high degree of control and precision~\cite{SwiMarCam99a}; (3) For sizable ranges of parameters, \FMF s appear insensitive to changes in one or more parameters. This is seen as near vertical or horizontal shaded lines in Fig.~\ref{fig:phase}. 

In particular, near the crossing points in Fig.~\ref{fig:phase}(d), a fluctuation in just one of $\mu_1$ and $\mu_2$ cannot remove the \FMF s. While not topological, this provides some robustness for the \FMF s in this system. In equilibrium a ``quadratic'' robustness exists~\cite{LeiFle12a} near $\mu_1=\mu_2=0$ with the Majorana fermion energy splitting proportional to $\mu_1\mu_2/2\Delta$. In the driven system, our numerical results~\cite{Note1} show the splitting is quadratic or linear reflecting the structures shown in Fig.~\ref{fig:phase}. We emphasize that the system proposed here can easily be re-tuned to stabilize the \FMF s by varying the manifold of control parameters.

\emph{Detection.}---The detection of Majorana fermions presents several challenges. They may be detected by spectroscopic measurements, such as tunneling spectra or conductance through leads that overlap with their wavefunctions. However, the topological character of Majorana fermions is difficult to ascertain in such experiments. In the case of \FMF s the dynamical nature of the states can be exploited for a clear identification. Two of us~\cite{KunSer13a} have shown before that the (zero-temperature) conductance $\sigma$ in a single-terminal (tunneling) or a symmetric two-terminal setup satisfies the Floquet sum rule with bias $V$,
$
\sum_{n\in\mathbb{Z}}\sigma(V+n\Omega) = {2e^2}/{h},
$
for $V=\varepsilon_0$ and/or $\varepsilon_\pi$ if and only if \FMF s are present at $\varepsilon=\varepsilon_0$ and/or $\varepsilon_\pi$. The topological nature of \FMF s protects the Floquet sum rule against disorder. In the current system of quantum dots, the Floquet sum rule can also be used to detect the \FMF s; however, the topological protection is lost.

\begin{figure}[t]
\centering
\includegraphics[width=0.49\textwidth]{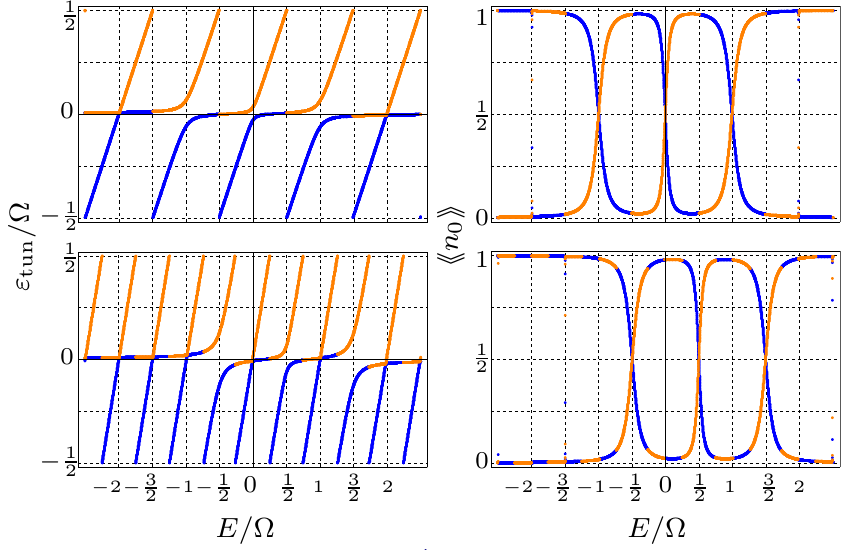}
\caption{(color online) Adiabatic detection of \FMF s. For the reduced system consisting of the probe dot $D_0$ and two degenerate \FMF s in the coupled dots $D_1$ and $D_2$, the tunneling quasienergy $\varepsilon_{\mathrm{tun}}$ (left), and the time-averaged probability $\bbraket{n_0}$ of finding an electron in $D_0$ (right), are shown as the energy $E$ of resonant level in $D_0$ varies. The top and bottom panels show, respectively, the results for \FMF\ at quasienergy $0$ and $\Omega/2$. The dark (blue) and light (orange) curves correspond to, respectively, lower and higher tunneling quasienergies. A simple profile of the tunneling amplitude $v(t) = e^{i\varepsilon t}(v_0 + v\cos \Omega t)$ is assumed with $v_0=0.2\Omega/2\pi, v=0.8\Omega/2\pi$. The probability follows a smooth curve that matches the crossing and anti-crossing points of the quasienergy spectrum.}
\label{fig:tun}
\end{figure}

A more suitable detection setup for the quantum-dot system is obtained by coupling a probe quantum dot, $D_0$, with a single resonant level at energy $E$ to one or both of the dots, as depicted in Fig.~\ref{fig:setup}. Since the fermion parity of the whole system is preserved, the reduced tunneling Hamiltonian takes the form~\cite{Fle11a}
\begin{equation}
H_{\mathrm{tun}}(t) \propto \left( \begin{array}{cc} 0 & v(t) \\ v(t)^* & E \end{array} \right),
\end{equation}
in either the odd or even parity subspace spanned by $|n_0n_{12}\rangle$ occupation basis, where $n_0=0,1$ is the occupation of $D_0$ and $n_{12}=0,1$ is the occupation of one pair of states in $D_1$ and $D_2$ with quasienergies $\pm\varepsilon$. For \FMF s at quasienergy $\varepsilon=\varepsilon_0$ and $\varepsilon_\pi$ one has $e^{i\varepsilon T}=e^{-i\varepsilon T}$ and the tunneling element, $v(t) = \langle\psi(t)|n_0\rangle = e^{i\varepsilon t} \langle\phi(t)|n_0\rangle $, is a periodic function with a period, respectively, $T$ and $2T$. Therefore, the adiabatic evolution of the system as $E$ is changed from $-\infty$ to $\infty$ results in characteristic switching transitions in the occupation (charge) of the probe dot at, respectively, $E=k\Omega$ and $(k+\frac12)\Omega$ with $k\in\mathbb{Z}$. These characteristic switchings can be used to detect \FMF s unambiguously. In Fig.~\ref{fig:tun}, a typical plot for the time-averaged value of the probe charge, $\bbraket{n_0}$, is shown for a simplified profile of $v(t)$. More details and results for the full three-dot system are given in the Supplemental Material~\cite{Note1}.

\emph{Discussion.}---The setup proposed here can be realized in the lab using well-established tools developed for fabricating and measuring quantum dots in order to create, detect, and manipulate \FMF s in a highly tunable fashion without a locally variable magnetic field. The tuning can be done instead in an all-electric circuit with great precision.  A potential challenge in creating this system is obtaining appreciable pairing amplitude between spin-filtered resonant levels of the two quantum dots through the superconductor bridge. For a singlet superconductor, this can be addressed in at least three ways: (i) by exploiting the spin-orbit interaction with a characteristic length smaller than the size of the quantum dots~\cite{SauSar12a,NadPriBer12b}; (ii) by applying opposite Zeeman fields on the two dots; or (iii) by applying a uniform external Zeeman field and field-tuning the quantum dot $g$-factors~\cite{SchPetJun11a} to achieve opposite signs on the two dots. 

A major obstacle in the clear detection of Majorana fermions is ruling out other mechanisms leading to bound state with similar spectroscopic signatures, e.g. disorder-induced Andreev bound states or Kondo resonances~\cite{LiuPotLaw12b,PikDahWim12b,LeeJiaAgu12a}. We have not considered the Kondo effect in response to the ac drive in our setting~\cite{LopAguPla01a}. However, we note that in static equilibrium the Kondo fixed point is unstable and the phase of the system is instead controlled by another fixed point induced by the Majorana fermions~\cite{CheBecBau13a}. It would be interesting to study the interplay between Kondo interaction and \FMF s. In this regard, an advantage of \FMF s over the static ones is the existence of $\varepsilon_{\pi}$ states that are distinct from $\varepsilon_0$ states and other non-dynamical bound states and whose presence can be clearly identified in transport or tunneling experiments we have described.

To summarize, we have discussed \FMF s in a double-dot system coupled through a superconducting and subject to ac gate potentials. The driving amplitude, frequency and the ac phase difference between the dots provide a large, highly tunable set of control knobs that make it possible to observe the signatures of \FMF s in a solid-state system and  to test their exotic properties. Extensions of the proposed architecture to many coupled dots can open a viable way to realize Floquet Majorana fermions in the lab as a platform for quantum information processing and non-local quantum entanglement.

This research is supported by the College of Arts and Sciences at Indiana University, Bloomington (A.K. and B.S.) and by CNSF, grant No.~10625420, and the FRFCU (Y.L. and F.Z.). Y.L would like to thank Yankui Wang for her initial collaboration and the support from China Scholarship Council.
\vspace{-5mm}

\bibliographystyle{physre}

\newpage~\newpage
\renewcommand{\thefigure}{S\arabic{figure}}
\setcounter{figure}{0}
\renewcommand{\theequation}{S\arabic{equation}}
\setcounter{equation}{0}

\section{Supplemental Material}

\subsection{Effective Hamiltonian in high frequency approximation}
\noindent
Here we sketch the derivation of the effective Hamiltonian leading to Eq. (6) of the main text. We compute the matrix elements of the Floquet Hamiltonian $H(t)-i\partial_t$ in the \textit{rotating wave} basis 
$|\{n_{i}\},m\rangle =\exp\left(im\Omega t-i\int_0^t V(s)ds \right)|\{n_{i}\}\rangle$, as

\begin{align}
&\langle \langle \{n_{i}\},m|H(t)-i\partial _{t}|\{n_{i}^{\prime
}\},m^{\prime }\rangle \rangle   \notag \\
&=\frac{1}{T}\int_{0}^{T}dt \langle \{n_{i}\},m|H(t)-i\partial
_{t}|\{n_{i}^{\prime }\},m^{\prime }\rangle   \notag \\
&=\delta _{mm^{\prime }}\left\langle \{n_{i}\}\right\vert H_{0}(\Delta
=\lambda =0)+m\Omega |\{n_{i}^{\prime }\}\rangle \notag \\
&+\frac{1}{T} \int_{0}^{T}dt\left\langle \{n_{i}\}\right\vert \Big[ V(t)e^{i(m^{\prime }-m)\Omega t}  \notag \\
&+ \Big( H_{\Delta}^{+}+Q_+H_{\Delta}^{-} +\frac{1}{2!}Q_+^{2}H_{\Delta}^{+}+\cdots \Big)
e^{i(m^{\prime }-m)\Omega t}  \notag \\
&+\Big(H_{\lambda }^{+}+ Q_- H_{\lambda }^{-}+\frac{1}{2!}Q_-^{2}H_{\lambda }^{+}+\cdots \Big)e^{i(m^{\prime }-m)\Omega t} \Big] |\{n_{i}^{\prime }\}\rangle , \label{LS}
\end{align}%
where, $H_{\Delta}^{\pm}=\Delta \left( d_{1}^{\dagger }d_{2}^{\dagger
}\pm d_{1}d_{2}\right) $, $H_{\lambda }^{\pm}=\lambda \left( d_{1}^{\dagger
}d_{2}\pm d_{1}d_{2}^{\dagger }\right) $ and $Q_{\pm} = i\frac{A_0}{\Omega}\left( \sin(\Omega t)d_1^{\dagger}d_1 \pm \sin(\Omega t+ \theta)d_2^{\dagger}d_2\right)$. 
Summing over the series in Eq.~(\ref{LS}), we have
\begin{align}
&\langle \{n_{i}\}|\Big( H_{\Delta}^{+}+Q_+H_{\Delta}^{-} +\frac{1}{2!}Q_+^{2}H_{\Delta}^{+}+\cdots \Big) |\{n_{i}\}\rangle\notag \\
& = \langle \{n_{i}\}| \left[ H_{\Delta}^+\cos \left( \frac{2A_{0}}{\Omega }\sin \left( \Omega t+\theta /2\right)
\cos \left( \theta /2\right) \right) \right.  \notag \\
&~ \left. + iH_{\Delta}^{-}\sin \left( \frac{2A_{0}}{\Omega }\sin \left( \Omega
t+\theta /2\right) \cos \left( \theta /2\right) \right) \right] |\{n_{i}\}\rangle
\end{align}%
and
\begin{align}
&\langle \{n_{i}\}| \Big(H_{\lambda }^{-}+ Q_- H_{\lambda }^{-}+\frac{1}{2!}Q_-^{2}H_{\lambda }^{+}+\cdots \Big) |\{n_{i}\}\rangle \notag \\
& = \langle \{n_{i}\}| \left[ H_{\lambda }^{+}\cos \left( \frac{2A_{0}}{\Omega }\cos (\Omega t+\theta /2)\sin
\left( \theta /2\right) \right) \right.    \notag \\
&~ \left. -iH_{\lambda }^{-}\sin \left( \frac{2A_{0}}{\Omega }\cos \left(
\Omega t+\theta /2\right) \sin \left( \theta /2\right) \right) \right]|\{n_{i}\}\rangle
\end{align}%
Noting $H^{(0)}$ corresponds to $m=m^{\prime }=0$, $H^{(1)}$ corresponds to $%
m^{\prime }-m=1$ and $H^{(-1)}$ corresponds to $m^{\prime }-m=-1$, we perform the
time integral, then the expressions of $H^{(0)}$, $H^{(1)}$ and $H^{(-1)}$
in Nambu space $\left( d_{1},d_{2},d_{1}^{\dagger }, d_{2}^{\dagger }\right) $ are
\begin{align}
H^{(0)}=\left(
\begin{array}{cccc}
\mu_{1} & J_{0}\left( \xi ^{\mathrm{s}}\right) \lambda  & 0 & J_{0}\left( \xi
^{\mathrm{c}}\right) \Delta  \\
J_{0}\left( \xi ^{\mathrm{s}}\right) \lambda  & \mu_{2} & -J_{0}\left( \xi ^{%
\mathrm{c}}\right) \Delta  & 0 \\
0 & -J_{0}\left( \xi ^{\mathrm{c}}\right) \Delta  & -\mu_{1} & -J_{0}\left(
\xi ^{\mathrm{s}}\right) \lambda  \\
J_{0}\left( \xi ^{\mathrm{c}}\right) \Delta  & 0 & -J_{0}\left( \xi ^{%
\mathrm{s}}\right) \lambda  & -\mu_{2}%
\end{array}%
\right),
\end{align}%
\begin{widetext}
\begin{align}
H^{(\pm1)}=\left(
\begin{array}{cccc}
A_{0}/2 & -ie^{\mp i\theta /2}J_{1}\left( \xi ^{\mathrm{s}}\right) \lambda  & 0
& \mp e^{\mp i\theta /2}J_{1}\left( \xi ^{\mathrm{c}}\right) \Delta  \\
ie^{\mp i\theta /2}J_{1}\left( \xi ^{\mathrm{s}}\right) \lambda  &
A_{0}e^{\mp i\theta }/2 & \pm e^{\mp i\theta /2}J_{1}\left( \xi ^{\mathrm{c}}\right)
\Delta  & 0 \\
0 & \mp e^{\mp i\theta /2}J_{1}\left( \xi ^{\mathrm{c}}\right) \Delta  & -A_{0}/2
& -ie^{\mp i\theta /2}J_{1}\left( \xi ^{\mathrm{s}}\right) \lambda  \\
\pm e^{\mp i\theta /2}J_{1}\left( \xi ^{\mathrm{c}}\right) \Delta  & 0 &
ie^{\mp i\theta /2}J_{1}\left( \xi ^{\mathrm{s}}\right) \lambda  &
-A_{0}e^{\mp i\theta }/2%
\end{array}%
\right),
\end{align}%
\end{widetext}
where $\xi^{\mathrm{c}}=2A_0\cos(\theta/2)/\Omega$ and $\xi^{\mathrm{s}}=2A_0\sin(\theta/2)/\Omega$, which gives the effective Hamiltonian%
\begin{align}
H_{\mathrm{eff}} &=H^{(0)}+\frac{\left[ H^{(-1)},H^{(1)}\right] }{\Omega } \notag \\
 &= \left(
\begin{array}{cccc}
\mu_{1} & \tilde{\lambda} & 0 & \tilde{\Delta} \\
\tilde{\lambda} & \mu_{2} & -\tilde{\Delta} & 0 \\
0 & -\tilde{\Delta} & -\mu_{1} & -\tilde{\lambda} \\
\tilde{\Delta} & 0 & -\tilde{\lambda} & -\mu_{2}%
\end{array}%
\right),
\end{align}%
with 
\begin{align}
\tilde{\Delta} &=\left[ J_{0}\left( \xi ^{\mathrm{c}}\right) -\xi ^{%
\mathrm{c}}J_{1}\left( \xi ^{\mathrm{c}}\right) \right] \Delta \\
\tilde{\lambda} &=\left[ J_{0}\left( \xi ^{\mathrm{s}}\right) -\xi ^{\mathrm{s}%
}J_{1}\left( \xi ^{\mathrm{s}}\right) \right] \lambda 
\end{align}
as mentioned in the
main text.

\begin{figure}[tb]
                \includegraphics[width=0.45\textwidth]{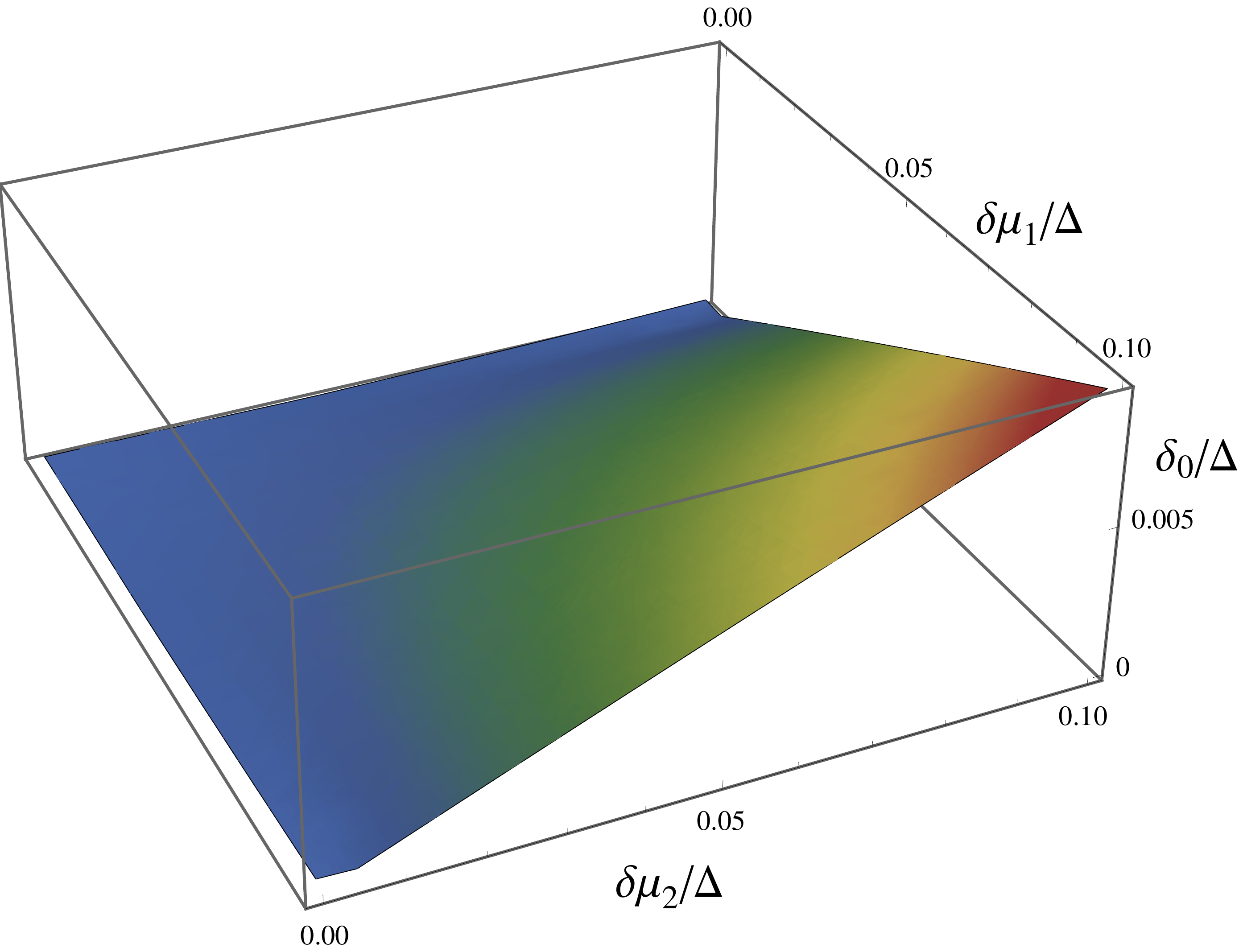}
        ~ 
                \includegraphics[width=0.45\textwidth]{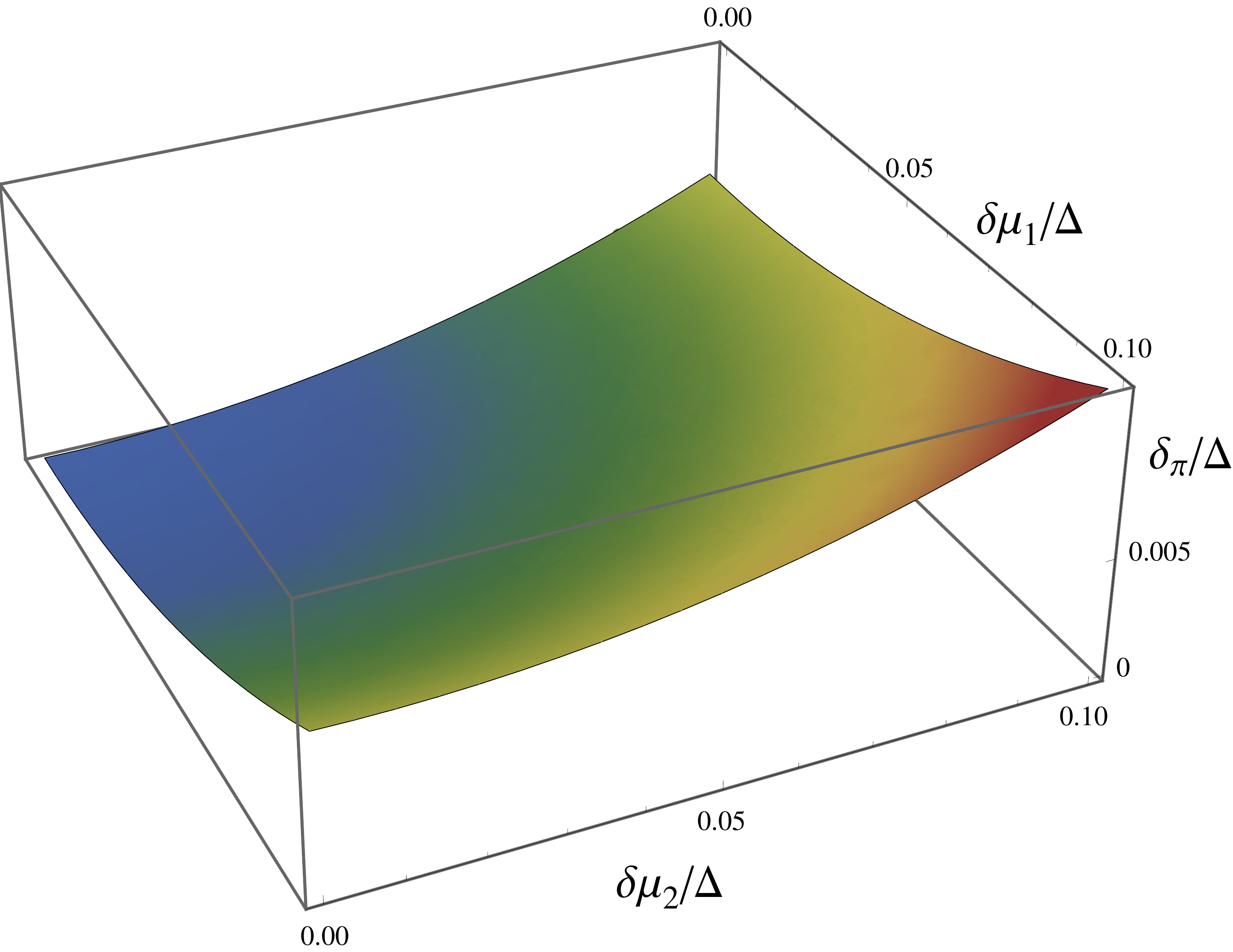}
 \begin{picture}(1,1)
\put(-250,320){(a)}
\put(-250,150){(b)}
\end{picture}
           \caption{The change $\delta_0 = |\varepsilon|T$ from $\varepsilon=\varepsilon_0 = 0$ (a) and $\delta_{\pi} = |\pi-\varepsilon T|$ from $\varepsilon=\varepsilon_{\pi} = \pi/T$  (b) with onsite energy deviations $\delta\mu_1$ and $\delta\mu_2$ around initial values $\mu_{10}=\mu_{20}=0$. For (a), a quadratic fit gives $\delta_0 \approx 10^{-4} + 0.96~\delta\mu_1\delta\mu_2/\Delta^2 - 0.012 (\delta\mu_1^2 + \delta\mu_2^2)/\Delta^2$ whereas, for (b), a quadratic fit gives $\delta_{\pi} \approx 0.095~\delta\mu_1\delta\mu_2/\Delta^2 - 0.64 (\delta\mu_1^2 + \delta\mu_2^2)/\Delta^2$. Other parameters used are $\lambda = 0.8\Delta,~A_0=1.5\Delta$. The driving frequencies $\Omega \approx 3.72 \Delta$ for (a) and $\Omega \approx 2.87 \Delta$ for (b) are fixed close to the the largest frequency for having a $\varepsilon_0$ and $\varepsilon_{\pi}$ states respectively for (a) and (b).\label{fig:S1}}
\end{figure}

\begin{figure}[tb]
                \includegraphics[width=0.45\textwidth]{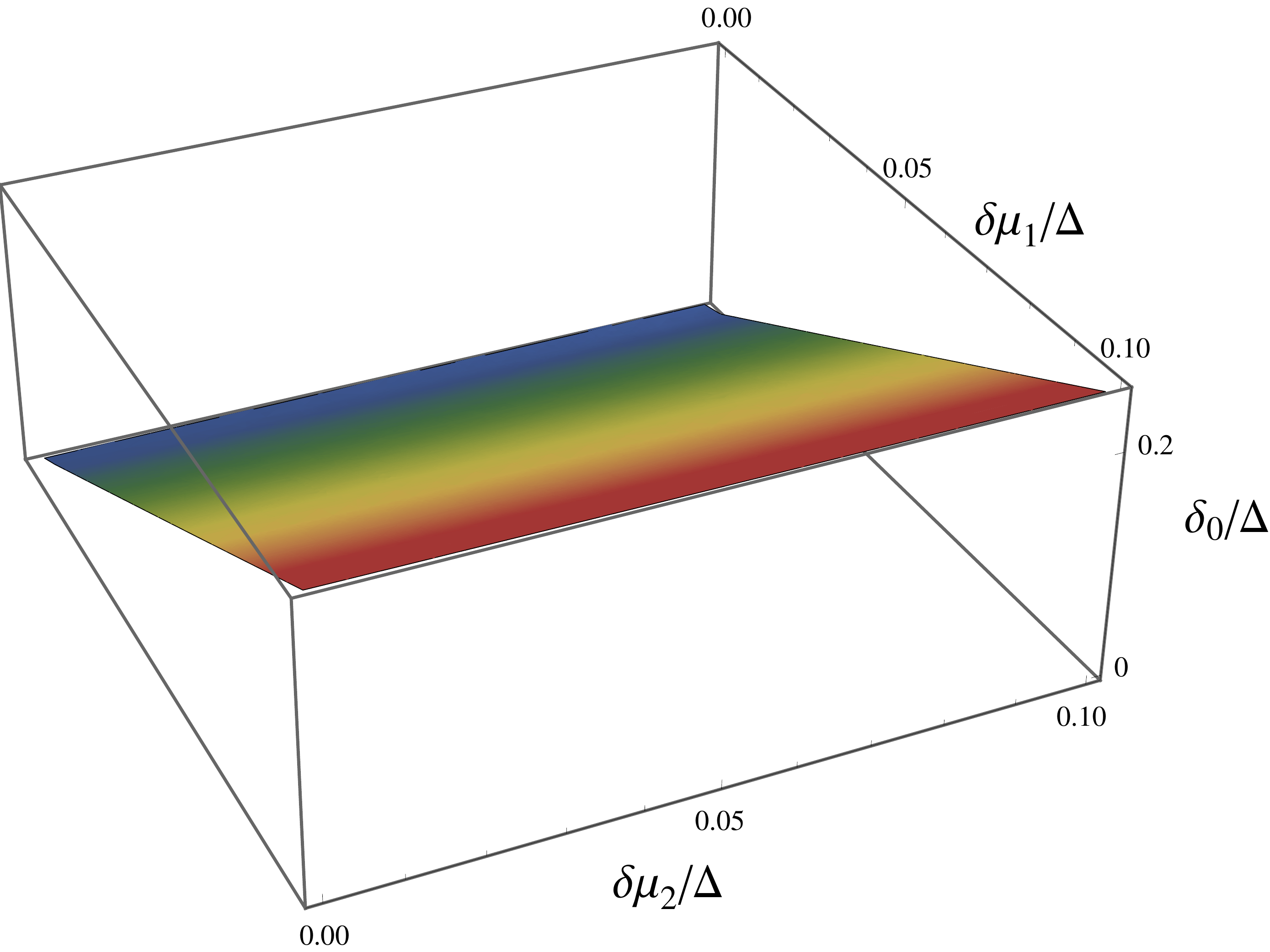}
                \includegraphics[width=0.45\textwidth]{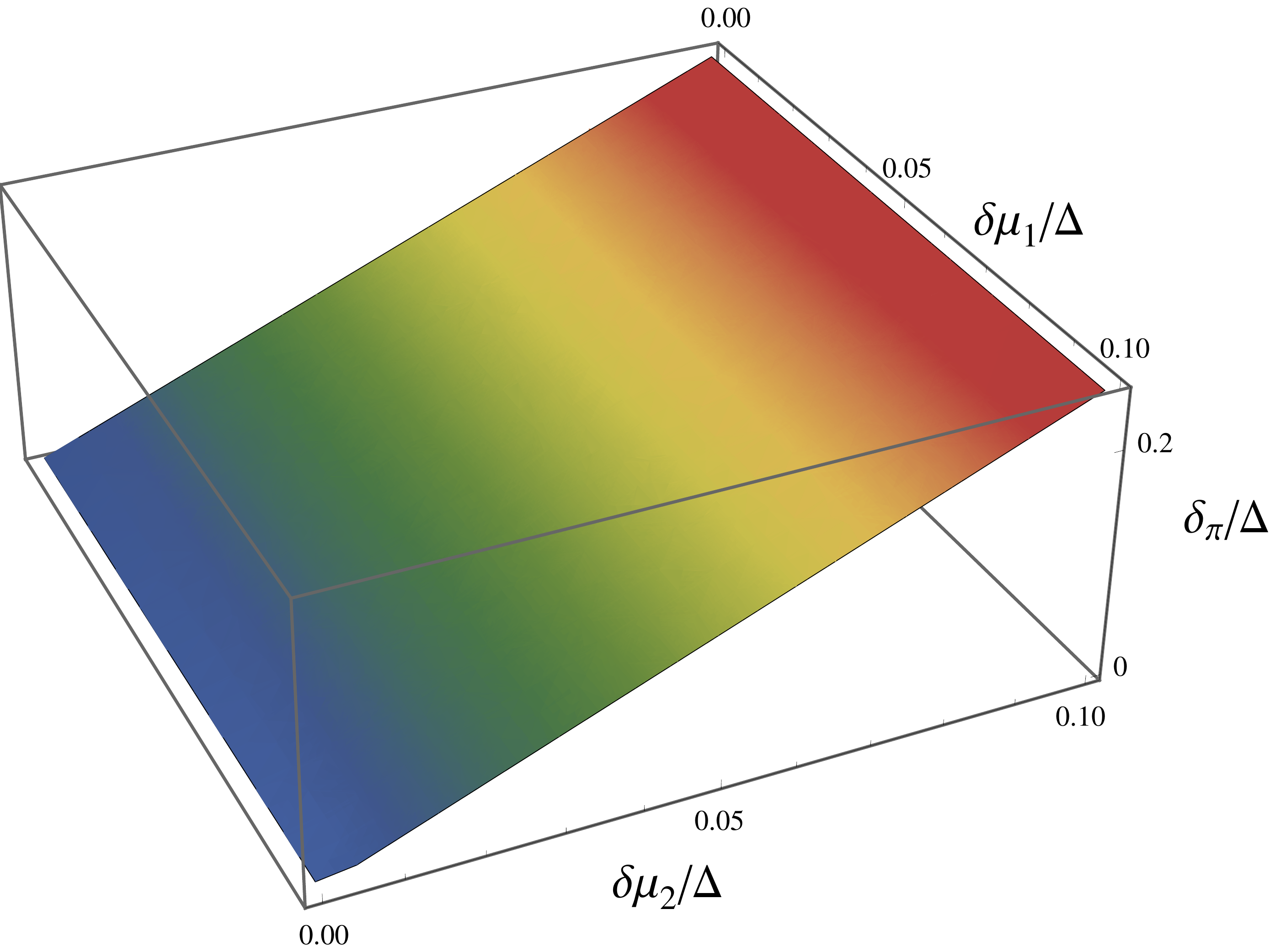}
\begin{picture}(1,1)
\put(-250,320){(a)}
\put(-250,150){(b)}
\end{picture}
        \caption{The change $\delta_0 = |\varepsilon|T$ from $\varepsilon=\varepsilon_0 = 0$ (a) and $\delta_{\pi} = |\pi-\varepsilon T|$ from $\varepsilon=\varepsilon_{\pi} = \pi/T$ (b) with onsite energies deviated from the condition of having $\varepsilon_0$ and $\varepsilon_{\pi}$ states by $\delta \mu_1 = \mu_1 - \mu_{10}$, $\delta \mu_2 = \mu_2 - \mu_{20}$. The initial parameters $\lambda = 0.8\Delta,~\mu_{10}=3.85,~\mu_{20}=1.05,~A_0=1.5\Delta$ and the driving frequency $\Omega \approx 2.15 \Delta$, the system has both the $\varepsilon_{0}$ and $\varepsilon_{\pi}$ states present. For (a), a quadratic fit gives $\delta_0 \approx 10^{-4} + 2.44~\delta\mu_1/\Delta - 0.093~ \delta\mu_1\delta\mu_2/\Delta^2 + 0.62~\delta\mu_1^2/\Delta^2 +  0.12~\delta\mu_2^2/\Delta^2$. Whereas, for (b), a quadratic fit gives $\delta_{\pi} \approx - 2.43~\delta\mu_2/\Delta - 0.068~ \delta\mu_1\delta\mu_2/\Delta^2 - 0.5~\delta\mu_2^2/\Delta^2$. }\label{fig:S2}
\end{figure}

\begin{figure*}[t]
\includegraphics[width=0.8\textwidth]{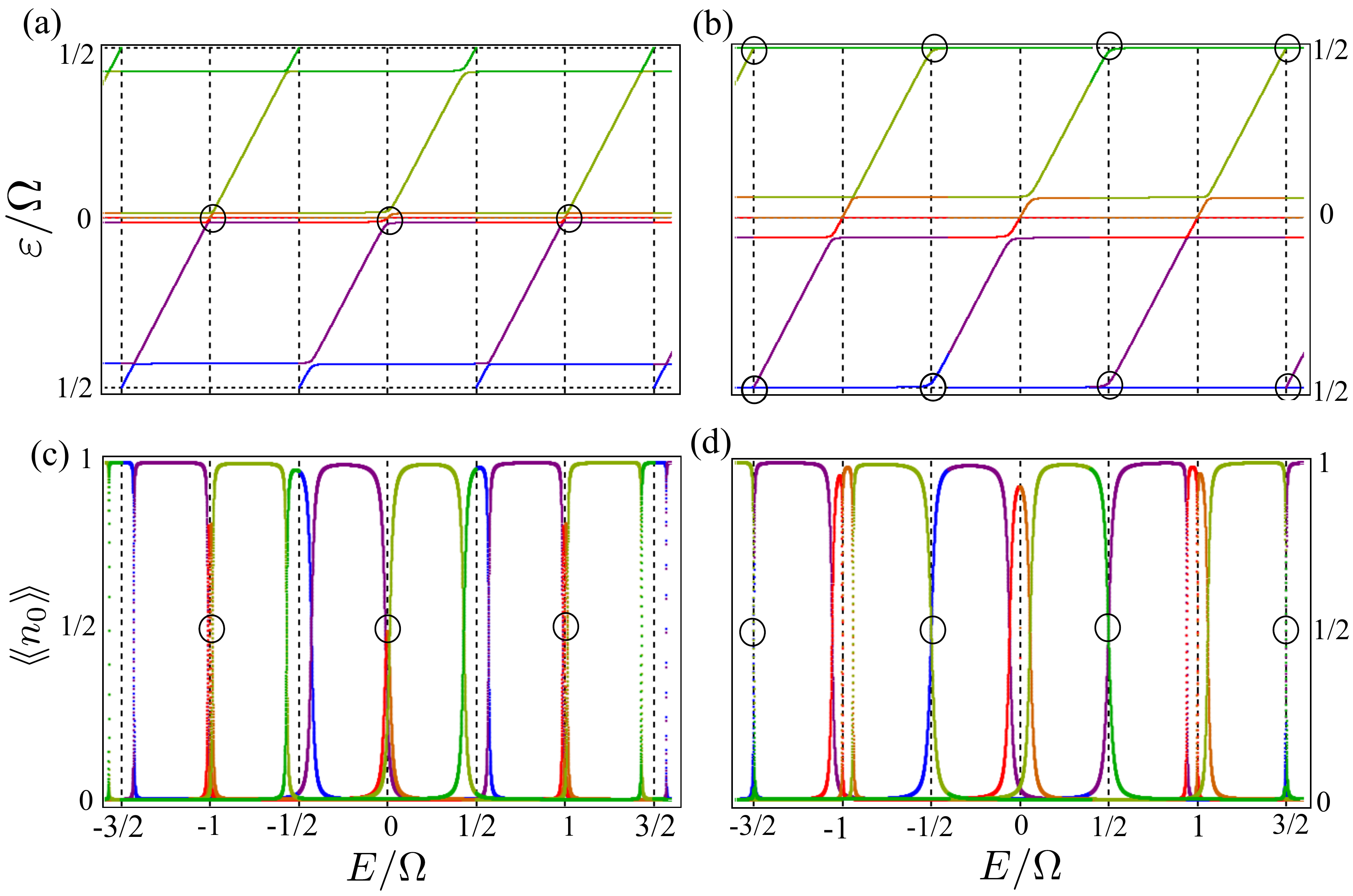}
\caption{Quasienergies of the 3-dot system with parameters the same as in Fig.~\ref{fig:S1} and $\Omega \approx 3.72 \Delta$ (a) and $\Omega \approx 2.87 \Delta$ (b), which are respectively the frequencies at which $\varepsilon_0$ and $\varepsilon_{\pi}$ states appear. The points of avoided crossing with $\varepsilon_0$ and $\varepsilon_{\pi}$ states are marked with circles.
 Switching transitions in the time-averaged charge $\mathinner{\langle\hspace{-0.75mm}\langle n_0\rangle\hspace{-0.75mm}\rangle}$ of the probe dot occurs as we cross $E = k\Omega$ (c) and $(n+\frac12)\Omega$ (d) in the presence of $\varepsilon_0$ (c) and $\varepsilon_{\pi}$ (d) Floquet Majorana states, respectively. Different colors represent  $\mathinner{\langle\hspace{-0.75mm}\langle n_0\rangle\hspace{-0.75mm}\rangle}$ in different steady states of the system. Other avoided crossings not marked in circles shows the presence of other quasienergy levels. The avoided crossings at multiples of $\Omega$ and odd multiples of $\Omega/2$ signify the presence of Floquet Majorana fermions.\label{figS2}}
\end{figure*}

\subsection{Robustness against onsite energies}
\noindent
The $\varepsilon_0$ or $\varepsilon_{\pi}$ states appear in the double-dot system for a fine-tuned set of parameters as we describe in the main text. Quasienergies will move away from $\varepsilon_0=0$ or $\varepsilon_{\pi}=\pi/T$ as we deviate from the fine-tuned point in the parameter space.
Here we present a numerical analysis of the stability of the Floquet Majorana fermions as a function of the deviations $\delta \mu_1 = \mu_1 - \mu_{10}$, $\delta \mu_2 = \mu_2 - \mu_{20}$ from the fine-tuned onsite energies $\mu_{10}$ and $\mu_{20}$. Our result shows, if $\mu_{10}=\mu_{20} = 0$, then quasienergies have quadratic dependence on deviations. However, for finite $\mu_{10}$ and $\mu_{20}$ the dependence is linear.

Fig.~\ref{fig:S1}(a) shows the change in quasienergy $\delta_0 = |\varepsilon|T$ away from $\varepsilon=\varepsilon_0 = 0$ as we increase both $\mu_1$ and $\mu_2$ from $\mu_{10}= \mu_{20} = 0$ by amounts $\delta\mu_1$ and $\delta\mu_2$ respectively. Numerically the best fit up to quadratic in $\delta\mu_1,~\delta\mu_2$ gives $\delta_0 \approx r_1 \delta\mu_1\delta\mu_2/\Delta^2 + r_2 (\delta\mu_1^2 + \delta\mu_2^2)/\Delta^2$ with $|r_2| \ll |r_1|$. This shows a ``quadratic protection'' against onsite energy fluctuations. The state around $\varepsilon_{\pi}$ also possesses a quadratic protection for $\mu_{10}=\mu_{20} = 0$. As shown in Fig.~\ref{fig:S1}(b), the change $\delta_{\pi} = | \pi - \varepsilon T|$ has a quadratic dependence as $\delta_{\pi} \approx r_1 \delta\mu_1\delta\mu_2/\Delta^2 + r_2 (\delta\mu_1^2 + \delta\mu_2^2)/\Delta^2$ with $|r_2| \gg |r_1|$.

For Fig.~\ref{fig:S2}, we have suitably chosen a point in parameter space for which $\mu_{10}, \mu_{20} > 0$ and possesses both the $\varepsilon_0$ and $\varepsilon_{\pi}$ states. The resulting change in quasienergies  are linear in $\delta\mu_1$ or $\delta\mu_2$, as long as they are small. But around this particular point, $\delta_0$ remains insensitive to the $\delta\mu_2$ deviation, whereas $\delta_{\pi}$ appears insensitive to $\delta\mu_1$.

\subsection{Detecting Floquet Majorana with a probe dot}
\noindent
We briefly discuss here an exact numerical setup for the 3-dot system we described in the main text. We add to the parent double-dot system a probe dot, $D_0$, with a spin-polarized energy-level $E$ which is coupled to one of the previous dots with a tunneling amplitude $v$. The probe dot is not periodically driven.

In the presence of Floquet Majorana states, the time averaged charge of the $D_0$ exhibits switching transitions as we change $E$ across $E = k\Omega$ and $(k+\frac12)\Omega$ (for integer $k$) for $\varepsilon_0$ and $\varepsilon_{\pi}$ states, respectively. There could be many other avoided crossings of quasienergy levels for a realistic system, but the switching exactly at energy multiples of $\Omega$ (or $\Omega/2$) uniquely determines the presence of $\varepsilon_0$ (or $\varepsilon_{\pi}$) mode. The direct numerical computation, as in Fig~\ref{figS2} matches our simplified model discussed in the main text in Fig. 3.

\end{document}